\documentclass[11pt]{article}
\usepackage{amssymb,latexsym,amsmath,amsbsy,amsthm}
\usepackage[dvips]{graphicx}
\usepackage{cite}
\usepackage{stmaryrd}  
\usepackage{arydshln}  
\usepackage{mathdots}  
\usepackage{young}  
\usepackage{pdfpages}
\newcommand{\myatop}[2]{\genfrac{}{}{0pt}{}{#1}{#2}}

\def\ss{\scriptstyle}
\def\nn{\nonumber}

\def\qdots{\mathinner{\mkern1mu\raise1pt\vbox{\kern7pt\hbox{.}}\mkern2mu \raise4pt\hbox{.}\mkern2mu\raise7pt\hbox{.}\mkern1mu}}
\def\Z{{\mathbb Z}}
\def\N{{\mathbb N}}

\def\osp{\mathfrak{osp}}

\def\beq{\begin{equation}}
\def\eeq{\end{equation}}

\begin{document}
\begin{center}
{\Large \bf
Partition functions and thermodynamic properties of paraboson and parafermion systems\\[5mm] }
{\bf N.I.~Stoilova}\footnote{Corresponding author}\footnote{E-mail: stoilova@inrne.bas.bg}\\[1mm] 
Institute for Nuclear Research and Nuclear Energy,\\ 
Boul.\ Tsarigradsko Chaussee 72, 1784 Sofia, Bulgaria\\[2mm] 
{\bf J.\ Van der Jeugt}\footnote{E-mail: Joris.VanderJeugt@UGent.be}\\[1mm]
Department of Applied Mathematics, Computer Science and Statistics, Ghent University,\\
Krijgslaan 281-S9, B-9000 Gent, Belgium.
\end{center}


\begin{abstract}
New formulas are given for the grand partition function of paraboson systems of order~$p$ with $n$ orbitals 
and parafermion systems of order~$p$ with $m$ orbitals.
These formulas allow the computation of statistical and thermodynamic functions for such systems.
We analyze and discuss the average number of particles on an orbital, and the average number of particles in the system.
For some special cases (identical orbital energies, or equidistant orbital energies) we can simplify the grand partition functions and describe thermodynamic properties in more detail. 
Some specific properties are also illustrated in plots of thermodynamic functions.
\end{abstract}

%

\setcounter{equation}{0}
\section{Introduction} \label{sec:Introduction}

Parastatistics  was introduced by Green in 1953~\cite{Green} as a generalization of ordinary 
Bose-Einstein and Fermi-Dirac statistics. The Fock space $V(p)$ of a system of $n$ pairs 
of paraboson operators $b_j^\pm$ $(j=1,\ldots,n)$ of order $p$ (referred to as an {\em $n$-paraboson system of order $p$}), 
where $p$ is a positive integer, is characterized by the trilinear relations~\cite{Green}
\begin{equation}
 [\{ b_{ j}^{\xi}, b_{ k}^{\eta}\} , b_{l}^{\epsilon}]= (\epsilon -\xi) \delta_{jl} b_{k}^{\eta} 
 +  (\epsilon -\eta) \delta_{kl}b_{j}^{\xi}, 
 \label{n-paraboson}
\end{equation}
and by the conditions
\begin{align}
& \langle 0|0\rangle=1, \qquad b_j^- |0\rangle = 0, \qquad (b_j^\pm)^\dagger = b_j^\mp,\nn\\
& \{b_j^-,b_k^+\} |0\rangle = p\,\delta_{jk}\, |0\rangle,
\label{pFockB}
\end{align}
with $j,k,l\in\{1,2,\ldots,n\}$ and $\eta, \epsilon, \xi \in\{+,-\}$ (to be interpreted as $+1$ and $-1$
in the algebraic expressions $\epsilon -\xi$ and $\epsilon -\eta$).

Similarly, the trilinear relations~\cite{Green}
\begin{equation}
[[f_{ j}^{\xi}, f_{ k}^{\eta}], f_{l}^{\epsilon}]=\frac 1 2
(\epsilon -\eta)^2
\delta_{kl} f_{j}^{\xi} -\frac 1 2  (\epsilon -\xi)^2
\delta_{jl}f_{k}^{\eta},  
\label{n-parafermion}
\end{equation}
where $j,k,l\in\{1,2,\ldots,m\}$, together with the conditions
\begin{align}
& \langle 0|0\rangle=1, \qquad f_j^- |0\rangle = 0, \qquad (f_j^\pm)^\dagger = f_j^\mp,\nn\\
& [f_j^-,f_k^+] |0\rangle = p\,\delta_{jk}\,|0\rangle,
\label{pFockF}
\end{align}
define an {\em $m$-parafermion system of order $p$}. The corresponding Fock space is denoted by $W(p)$.

Using~\eqref{pFockB} and \eqref{pFockF} the number operators $N_i$ for the paraboson and parafermion systems 
are defined by~\cite{GreenbergMessiah65}
\begin{align}
& \hat N_i=\frac{1}{2}\{b_i^+,b_i^-\}-\frac{p}{2}, \label{NoperB}\\
& \hat N_i=\frac{1}{2}[f_i^+,f_i^-]+\frac{p}{2}.
\label{NoperF}
\end{align}

A quantum state of the paraboson system is a vector $\psi$ of the Fock space $V(p)$ that is a common eigenvector of the 
number operators~\eqref{NoperB},
\beq
\hat N_i\;\psi = N_i\; \psi \qquad (i=1,\ldots,n).
\label{psi}
\eeq
The non-negative integers $N_i$ are interpreted as occupation numbers, describing the number of ``particles on orbital~$i$'' 
when the system is in the state $\psi$.
There is no restriction on these occupation numbers: $N_i\in\N=\{0,1,2\ldots\}$.
When $p=1$ there is (up to a normalization factor) a unique state with occupation numbers $(N_1,N_2,\ldots,N_n)$ for any
element of $\N^n$. 
In fact, for $p=1$ the paraboson statistics reduces to the ordinary Bose-Einstein statistics.
For $p\geq 2$ there are in general multiple states with the same occupation numbers, and the description of all states is non-trivial.

For parafermion systems of order~$p$, the situation is similar but now the occupation numbers in~\eqref{psi} satisfy
$N_i\in\{0,1,\ldots,p\}$. This is known as the {\it Pauli principle} for the parafermion statistics of order $p$: 
no more than $p$ particles can be on the same orbital.
For  $p=1$ parafermion statistics is reduced to the ordinary Fermi-Dirac statistics. 

Analyzing the structure of the Fock spaces $V(p)$ and $W(p)$ is a major problem.
In principle, the paraboson and parafermion Fock spaces can be constructed by means of the so called Green ansatz~\cite{Green}. 
This reduces to finding a basis of an irreducible constituent of a $p$-fold tensor product (see Theorem in~\cite{GreenbergMessiah65} on p. 1158), which turns out to be difficult.
More than five decades after para\-statistics was introduced the irreducible representations 
of the paraboson~\eqref{n-paraboson} and parafermion~\eqref{n-parafermion} trilinear relations 
with an unique vacuum~\eqref{pFockB} and~\eqref{pFockF} were 
constructed in~\cite{paraboson} and \cite{parafermion} using group theoretical techniques.

In the recent years parastatistics became again a field of increasing interest. 
For example, parabosons and parafermions were considered as 
candidates for the particles of dark matter/dark energy~\cite{Nelson, KY}.  
Furthermore, quantum simulation of parabosons~\cite{Alderete2017} and parafermions~\cite{Alderete2018} 
were proposed, thus providing a tool of potential use of paraparticles in designing quantum information systems. 
In order to clarify these possible applications of paraparticles they must also be investigated 
from the point of view of statistical thermodynamics. 

For this purpose, one should have proper forms of the grand partition function (GPF) of paraboson and parafermion systems.
Although these GPF's are known in some expanded form, one needs their expressions as generating functions in order to analyze statistical properties.
Such expressions are known only in some special cases (such as $p=1$ and $p\geq n$ for the $n$-paraboson system of order~$p$).
In the present paper, we shall deduce proper forms of the grand partition function of paraboson and parafermion systems for all cases (i.e.\ for all values of~$p$).
This follows from recent work on the characters of the Fock spaces $V(p)$ and $W(p)$~\cite{paraboson,parafermion}.
Using these GPF's, we can investigate some thermodynamic properties of paraboson and parafermion statistics, such as 
the average number of particles on an orbital, or the total number of particles in the system. 

The structure of the paper is as follows. 
In Section~\ref{sec:A} we briefly summarize some of the mathematical notions that are needed in this paper. These concern partitions, Young diagrams and symmetric functions.
Section~\ref{sec:B} deals with statistical and thermodynamic properties of $n$-paraboson systems of order~$p$.
Due to a new character formula from representation theory, one has a general formula for the grand partition function for such systems.
We discuss this formula and some related statistical functions.
In particular, we show how it simplifies for two special cases (identical energy levels, and equidistant energy levels). 
For these two special cases, relevant thermodynamic properties are illustrated by plotting some distribution functions.
In Section~\ref{sec:C} the same analysis is performed for an $m$-parafermion system of order~$p$. 
Also here, a new formula for the grand partition function can be used to derive statistical and thermodynamic properties.
Due to these new formulas, we can derive a remarkable connection between the GPF's for paraboson and parafermion systems, 
at least in one of the special cases (identical energy levels). 
This is given in Section~\ref{sec:D}, where we also give some new alternative formulas for parastatistic systems with $p=2$ or $p=3$.
The paper ends with a short summary and some conclusions.


\section{Preliminaries and definitions}
\setcounter{equation}{0} \label{sec:A}

We need some basic notions on partitions and symmetric functions, see~\cite{Mac} as a standard reference.
A partition $\lambda=(\lambda_1,\lambda_2,\ldots,\lambda_n)$ of weight $|\lambda|$ and length $\ell(\lambda)\leq n$
is a sequence of non-negative integers satisfying the condition $\lambda_1\geq\lambda_2\geq\cdots\geq\lambda_n$, such that their
sum is $|\lambda|$, and $\lambda_i>0$ if and only if $i\leq \ell(\lambda)$. 
To each such partition there corresponds a Young diagram $F^\lambda$ consisting of $|\lambda|$ boxes 
$z=(i,j)\in \Z^2$, arranged in $\ell(\lambda)$ left-adjusted
rows of lengths $\lambda_i$ for $i=1,2,\ldots,\ell(\lambda)$. 
The first coordinate $i$ (the row index)
increases as one goes downwards, and the second coordinate $j$ (the column index) increases as one goes from left to right.
For example, the Young diagram of $\lambda=(5,4,4,2)$ is given by
\[
\begin{Young}
&&&&\cr
&&&\cr
&&&\cr
&\cr
\end{Young}
\]
The conjugate partition $\lambda'$ corresponds to the Young diagram of $\lambda$ reflected about the main diagonal. 
In other words, $\lambda_j'$ is the length of column~$j$ of $F^\lambda$. For the above example, $\lambda'=(4,4,3,3,1)$.
For each box $z=(i,j)$ in $F^\lambda$ one defines the contents $c(z)$ and the hook length $h(z)$:
\begin{equation}
c(z)=j-i, \qquad 
h(z)=\lambda_i +\lambda_j'-i-j+1.
\end{equation}
An important notion is the Frobenius notation~\cite{Mac} of a partition $\lambda$. If $F^\lambda$ has $r=r(\lambda)$ boxes 
on the main diagonal, 
$r$ is said to be the rank of $\lambda$. 
In the above example, $r=3$, denoted by crosses in the diagonal boxes:
\[
\begin{Young}
$\times$&&&&\cr
&$\times$&&\cr
&&$\times$&\cr
&\cr
\end{Young}
\]
The arm lengths $a_k=\lambda_k-k$ and leg lengths $b_k=\lambda_k'-k$ ($k=1,\ldots,r$) refer to the remaining boxes to the right or below the $k$th diagonal box, where $a_1>a_2>\cdots>a_r\geq 0$ and $b_1>b_2>\cdots>b_r\geq 0$.
The Frobenius notation of $\lambda$ is then
\[
\lambda=\left( \myatop{a_1\ a_2 \cdots a_r}{b_1\ b_2 \cdots b_r} \right).
\]
Note that for our example we have $\lambda = \left( \myatop{4\, 2\, 1}{3\, 2\, 0} \right)$.

Partitions are used to label symmetric polynomials in $n$ independent variables $\boldsymbol x=(x_1,x_2,\ldots,x_n)$.
Of particular importance are the Schur polynomials~\cite{Mac} or $S$-functions $s_\lambda(\boldsymbol x)$.
There are various ways to define Schur polynomials. For a partition $\lambda$ with $\ell(\lambda)\leq n$, one has
\begin{equation}
s_\lambda(\boldsymbol x) = \frac{\det (x_i^{\lambda_j+n-j})_{1\leq i,j\leq n}}{\det (x_i^{n-j})_{1\leq i,j\leq n}} .
\label{schur}
\end{equation}
If $\ell(\lambda)>n$, one puts $s_\lambda(\boldsymbol x)=0$. Clearly, for the zero partition $\lambda=(0)$ (which is the only partition with 
Frobenius rank $r=0$) one has $s_{(0)}(\boldsymbol x)=1$.

For symmetric polynomials in general, and for Schur polynomials in particular, there are many interesting series or generating functions.
A famous example is
\begin{equation}
\sum_{\lambda} s_\lambda (x_1,\ldots,x_n) = \sum_{\lambda} s_\lambda (\boldsymbol x) =
\frac{1}{\prod_{i=1}^n(1-x_i)\prod_{1\leq j<k\leq n}(1-x_jx_k)},
\label{SumSchur}
\end{equation}
due to Cauchy and Littlewood~\cite{Little}. 
In the left hand side of~\eqref{SumSchur}, the (infinite) sum runs over the set of all possible partitions $\lambda$.
The right hand side can be interpreted as a generating function.

When the variables $(x_1,x_2,\ldots,x_n)$ take a special form, the Schur polynomial simplifies.
We need two such simplifications in this paper.
The first case is when all $x_i$'s are equal, i.e.\ $x_1=x_2=\cdots=x_n=x$. 
Then (see~\cite{Mac}, p.~45)
\begin{equation}
s_\lambda (\underbrace{x,\cdots,x}_{\mbox{$n$ times}}) = 
\Big(\prod_{(i,j)\in \lambda}\frac{n+j-i}{\lambda_i+\lambda_j'-i-j+1}\Big) x^{|\lambda|}=
\Big( \prod_{z\in \lambda}\frac{n+c(z)}{h(z)}\Big) x^{|\lambda|}. 
\label{SumSchurx}
\end{equation}
The second case is when the consecutive ratios of the $x_i$'s is constant, i.e.\ for
$x_1=x$, $x_2=qx$, $\cdots$, $x_i=q^{i-1}x$, $\cdots$, $x_n=q^{n-1}x$ (see~\cite{Mac}, p.~44)
\begin{equation}
s_\lambda (x, qx,\cdots, q^{n-1}x) = q^{n(\lambda)} 
\Big(\prod_{z\in \lambda}\frac{1-q^{n+c(z)}}{1-q^{h(z)}}\Big) x^{|\lambda|}, 
\label{SumSchurqx}
\end{equation}
where
\[ 
n(\lambda)=\sum_{i\geq 1} (i-1)\lambda_i.
\]

\section{Thermodymanic properties of paraboson systems}
\setcounter{equation}{0} \label{sec:B}

We will assume that the $n$-paraboson system of order $p$ governed by~\eqref{n-paraboson}--\eqref{pFockB} 
is in a thermal and diffusive
contact and in a thermal and diffusive equilibrium with a much
bigger reservoir. We denote by $\tau= k_BT$ its (fundamental) temperature,
by $\mu_i$ the chemical potential and by $\epsilon_i$ the energy
for the particles on orbital~$i$. 
Then the grand partition function of the system is
the sum of the Gibbs factors with respect to all states of the system~\cite{Kittel}.
From the results in~\cite{GreenbergMessiah65, Hartle1,Hartle2,Hartle3} it follows that the structure of the GPF
for a system corresponding to quantum statistics based on the permutation group and defined in terms 
of an algebra of creation $a_i^+$ and annihilation $a_i^-$ ($i=1,2,\cdots, n$) operators, is in general given by
\begin{equation}
Z(x_1, x_2, \cdots, x_n ) = {\sum_\lambda}^\prime \ s_\lambda (x_1, x_2, \cdots, x_n), \label{GPF_general}
\end{equation}
where the prime on the sum on the RHS of~\eqref{GPF_general} indicates the possible restrictions on the partitions 
$\lambda$  and
\begin{equation}
x_i=\exp (\frac{\mu_i-\epsilon_i}{\tau}) \label{xi}.
\end{equation}
The GPF corresponding to a particular quantum statistics based on the permutation group is 
obtained by specifying the restrictions in the RHS of~\eqref{GPF_general}.

For the $n$-paraboson system of order $p$ the sum in~\eqref{GPF_general} is for those partitions whose length 
$l(\lambda)$ is less than or equal to $p$
\begin{equation}
Z_{\rm pB}(n,p) = \sum_{\lambda, \; l(\lambda)\leq p} s_\lambda (x_1, x_2, \cdots, x_n). \label{pBGPF_general}
\end{equation}
Surprisingly, proper forms of $Z_{\rm pB}(n,p)$ as a generating function were not known so far, apart from the 
trivial case $p=1$ and $p\geq n$~\cite{Suranyi, Hama, Chaturvedi96, Chaturvedi97a,  Chaturvedi97b, Meljanac},
even though the cases $1<p<n$ are the most interesting for investigating thermodynamic properties of the system.

GPF's of the form~\eqref{pBGPF_general} consist of a sum over all possible states of the system (including multiplicities).
Hence they are equal to the character of the corresponding Fock space $V(p)$.
These spaces $V(p)$ are representations of $\osp(1|2n)$, and their characters have been determined and analyzed in~\cite{paraboson}.
Following that analysis, we are able to deduce generating functions of $Z_{\rm pB}(n,p)$ for the missing cases.

\subsection{General GPF for $n$-paraboson systems}
\label{3.1}

Using the character formulas for the paraboson Fock spaces of order $p$ given in~\cite{paraboson},
one can rewrite~\eqref{pBGPF_general} in the following form:
\begin{equation}
Z_{\rm pB}(n, p) = \frac{E_{(n;0,p)}}{\prod_i(1-x_i) \prod_{j<k}(1-x_jx_k) },
\label{GPF}
\end{equation}
for any positive integer~$p$.
Herein, $E_{(n;0,p)}$ is a {\em polynomial} in the variables $x_i$, hence~\eqref{GPF} is a proper generating function for the GPF.
The explicit form of $E_{(n;0,p)}$ is given by
\begin{equation}
E_{(n;0,p)} \equiv E_{(n;0,p)}(x_1,\ldots,x_n) =\sum_{\eta} (-1)^{c_\eta} s_\eta (\boldsymbol{x}).
\label{E}
\end{equation}
Herein, the sum runs over all partitions $\eta$ of the form
\begin{equation}
\eta = \left(
\begin{array}{cccc}
a_1 & a_2 & \cdots & a_r \\
a_1+p & a_2+p & \cdots & a_r+p
\end{array} \right)
\label{eta}
\end{equation}
in Frobenius notation, and 
\begin{equation}
c_\eta= a_1+a_2+\cdots+a_r + r.
\label{c_eta}
\end{equation}
The subscript in $E_{(n;0,p)}$ refers to the fact that $0$ is added at the top and $p$ added at the bottom of 
self-conjugate partitions $\left(\begin{array}{ccc}
a_1 &  \cdots & a_r \\ a_1 & \cdots & a_r \end{array} \right)$ in~\eqref{eta}.
The number of variables $x_1,\ldots, x_n$ is finite, thus the Schur polynomial $s_\eta(\boldsymbol{x})$ cancels when the length $\ell(\eta)>n$,
and therefore the expression $E_{(n;0,p)}$ is also finite. 
As a simple example for $n=4$ and $p=2$, 
\beq
E_{(4;0,2)} =  1- s_{(1,1,1)}+ s_{(2,1,1,1)}- s_{(2,2,2,2)},
\label{exE}
\eeq
with all Schur polynomials in four variables $(x_1,x_2,x_3,x_4)$.
For general $n$ and $p$ it is not so hard to compute $E_{(n;0,p)}$, and one ends up with a finite alternating
sum of Schur functions like in~\eqref{exE}.

The finite sums $E_{(n;0,p)}$ simplify in some special cases, see~\cite{paraboson}:
\begin{align}
& E_{(n;0,1)} = \prod_{1\leq j<k\leq n} (1-x_j x_k), \label{E1}\\
& E_{(n;0,n-1)}  = 1-x_1x_2\cdots x_n, \label{En-1} \\
& E_{(n;0,p)}  = 1, \hbox{ for } p\geq n.\label{E0p} 
\end{align}
The first special case $p=1$ corresponds to the grand partition function for ordinary bosons:
\begin{equation}
Z_{\rm pB}(n, 1) = \frac{1 }{\prod_i(1-x_i) }. \label{GPF1}
\end{equation}
The third special case yields
\begin{equation}
Z_{\rm pB}(n, p) = \frac{1}{\prod_i(1-x_i) \prod_{j<k}(1-x_jx_k) } \qquad\hbox{for } p\geq n
\label{GPFn}
\end{equation}
and this is also one of the known cases~\cite{Chaturvedi96}.
The second special case $p=n-1$ gives a new interesting expression
\begin{equation}
Z_{\rm pB}(n, n-1) = \frac{(1-x_1x_2\cdots x_n)}{\prod_i(1-x_i) \prod_{j<k}(1-x_jx_k) }.
\label{GPFn-1}
\end{equation}

When a grand partition function $Z=Z(x_1,\ldots,x_n)$ for a system is known, 
many thermodynamic functions such as entropy, heat capacity, particle distributions $\ldots$ can be computed from $Z$~\cite{Kittel}.
Here, we will be dealing with  
the average number of particles $\bar{l}_i$ on the $i$th orbital and the average number of 
particles $\bar N$ in the system, given by the following general formulas~\cite{Kittel}:
\begin{align}
& \bar{l}_i = x_i \partial_{x_i} (\ln Z ), \label{li} \\
& \bar{N} = \sum_{i=1}^n x_i \partial_{x_i} (\ln Z ) \label{av}.
\end{align}
Applying these general formulas for the $n$-paraboson system of order~$p$, one finds the following expressions:
\begin{itemize}
\item For $p\geq n$~\eqref{GPFn} and \eqref{li}--\eqref{av} give
\begin{equation}
\bar{l}_i = \frac{x_i}{1-x_i}+\sum_{\myatop{\ss j=1}{\ss j\neq i}}^n \frac{x_ix_j}{1-x_ix_j}, \quad 
\bar{N} = \sum_{i=1}^n \frac{x_i}{1-x_i}+2\sum_{ j<i}^n \frac{x_ix_j}{1-x_ix_j}.
\label{av_p>n-1}
\end{equation}
\item For $p=1$, the case~\eqref{GPF1} of ordinary bosons:
\begin{equation}
\bar{l}_i = \frac{x_i}{1-x_i}, \quad 
\bar{N} = \sum_{i=1}^n \frac{x_i}{1-x_i}.
\label{av_p=1}
\end{equation}
\item
For $p=n-1$, using~\eqref{GPFn-1}:
\begin{align}
&\bar{l}_i = \frac{x_i}{1-x_i}+\sum_{\myatop{\ss j=1}{\ss j\neq i}}^n \frac{x_ix_j}{1-x_ix_j}-\frac{x_1x_2\cdots x_n}{1-x_1x_2\cdots x_n},\nn \\
&\bar{N} = \sum_{i=1}^n \frac{x_i}{1-x_i}+2\sum_{ j<i}^n \frac{x_ix_j}{1-x_ix_j}-\frac{nx_1x_2\cdots x_n}{1-x_1x_2\cdots x_n}.
\label{av_p=n-1}
\end{align}
\end{itemize} 
For the remaining cases $p\in\{2, 3, \ldots,n-2\}$ not many simplifications take place, and the expressions 
still contain $E_{(n;0,p)}$:
\begin{align}
& \bar{l}_i = \frac{x_i}{1-x_i}+\sum_{\myatop{\ss j=1}{\ss j\neq i}}^n \frac{x_ix_j}{1-x_ix_j}
+\frac{x_i}{E_{(n;0,p)}}\frac{\partial E_{(n;0,p)}}{\partial{x_i}}, \nn\\
& \bar{N} = \sum_{i=1}^n \frac{x_i}{1-x_i}+2\sum_{ j<i}^n \frac{x_ix_j}{1-x_ix_j}+
\sum_{i=1}^n\frac{x_i}{E_{(n;0,p)}}\frac{\partial E_{(n;0,p)}}{\partial{x_i}}.
\label{av_p}
\end{align}

In the following subsections we shall consider two special cases where the GPF and the thermodynamic functions simplify.

\subsection{$n$-paraboson systems with identical energy levels}
\label{3.2}

Let us consider the case when all orbitals have the same energy, and let
us furthermore assume that they also have the same chemical
potential, i.e.\ $\mu_1=\mu_2=\cdots=\mu_n=\mu$. Therefore 
$x_1=x_2=\cdots =x_n =x$, with
\begin{equation}
x=\exp (\frac{\mu-\epsilon}{\tau}).
\end{equation}
The thermodynamic functions for this example follow from the above
considered formulas, under the specification
$x_i=x$ ($i=1,\ldots,n$).
It will be convenient to use the notation $\breve{Z}$ and $\breve{E}_{(n;0,p)}$ for this:
\beq
\breve{Z}_{\rm pB}(n,p) = Z_{\rm pB}(n,p)|_{x_i=x}, \qquad \breve{E}_{(n;0,p)}={E}_{(n;0,p)} (x,x,\ldots,x).
\eeq
Observe that thanks to~\eqref{SumSchurx}, the expressions for $\breve{E}_{(n;0,p)}$ simplify drastically.
For instance, following the earlier example~\eqref{exE}, one has
\beq
\breve{E}_{(4;0,2)} =  1 -4 x^3+4x^5-x^8 ,
\label{exEx}
\eeq
and thus
\beq
\breve{Z}_{\rm pB}(4,2)=\frac{1 -4 x^3+4x^5-x^8}{(1-x)^4 (1-x^2)^6}.
\eeq

As more general examples, we mention:
\begin{itemize}
\item For $p\geq n$:
\begin{equation}
\breve{Z}_{\rm pB}(n, p) = \frac{1}{(1-x)^n (1-x^2)^{n(n-1)/2} }, 
\label{GPFx}
\end{equation}
\begin{equation}
\bar{l}_i = \frac{x}{1-x}+ \frac{(n-1)x^2}{1-x^2}, \qquad 
\bar{N} = \frac{nx(1+nx)}{1-x^2}.
\label{av_p>n-1x}
\end{equation}
\item For $p=1$ (bosons):
\begin{equation}
\breve{Z}_{\rm pB}(n, 1) = \frac{1}{(1-x)^n }, \qquad \bar{l}_i = \frac{x}{1-x}, \qquad 
\bar{N} = \frac{nx}{1-x}.
\label{av_p=1x}
\end{equation}
\item For $p=n-1$:
\begin{equation}
\breve{Z}_{\rm pB}(n, n-1) = \frac{1-x^n}{(1-x)^n(1-x^2)^{n(n-1)/2} },
\end{equation}
\begin{equation}
\bar{l}_i = \frac{x(1+nx)}{1-x^2}-\frac{x^n}{1-x^n}, \qquad 
\bar{N} = \frac{nx(1+nx)}{1-x^2}-\frac{nx^n}{1-x^n}.
\label{av_p=n-1x}
\end{equation}
\end{itemize}

Let us consider in more detail the dependence on the energy of the average number of particles in the system, 
i.e.\ the distribution function.
We will denote $\bar{N}$ as $\bar{N}(n,p)$ to specify that we are dealing with an $n$-paraboson system of order~$p$.
In particular we want to examine the behavior of $\bar{N}(n,p)$ for different values of~$p$.
In such a context, it is common to take as variable 
\begin{equation}
y=\frac{\epsilon-\mu}{\tau},
\label{y}
\end{equation} 
which is the orbital energy in units of $\tau$; hence we have $x=\exp(-y)$.
In Figure~1 we plot $\bar{N}(n=4,p)$ for $p=1,2,3,4,5,\ldots$. 
The lowest curve is for $p=1$ and yields the Bose distribution function.
The curve for $p=2$ is higher, then follow the curves for $p=3$ and $p=4$. 
The curves for $p>4$ coincide with that of $p=4$, following~\eqref{av_p>n-1x}.  
So compared to an $n$-boson system, for a fixed energy the average number of particles for an $n$-paraboson system with $p>1$ is higher and increases as $p$ increases.

\subsection{$n$-paraboson systems with equidistant energy levels}
\label{3.3}

The next interesting case to consider is when the orbitals have equidistant energies $\epsilon_i$.
Let the gap between the different energy
levels be $\Delta >0$.
Then
\begin{equation}
\epsilon_i=\epsilon_1+(i-1)\Delta \qquad (i=1,2,\ldots,n).
\label{eni}
\end{equation}
We also assume that $\mu_1=\mu_2=\cdots=\mu_n=\mu$.
Under these conditions the different orbitals correspond to different
energy levels.
According to notation~(\ref{xi}), we have
\begin{equation}
x_i= \exp \left( \frac{\mu-\epsilon_i}{ \tau} \right)
= \exp \left( \frac{\mu-\epsilon_1}{ \tau} \right)
\exp \left(- \frac{\Delta}{ \tau} \right)^{i-1}
= x q^{i-1},
\label{xiequid}
\end{equation}
where we have used 
\begin{equation}
x=x_1=\exp \left(\frac{\mu-\epsilon_1}{\tau} \right)
\qquad\hbox{ and }\qquad
q=\exp \left( -\frac{\Delta}{\tau}\right).
\label{xq}
\end{equation}
Under the specialization~\eqref{xiequid} and~\eqref{xq} the GPF's and the average number of particles 
$\bar{l}_i$ on the $i$th orbital can again be simplified.
We will now use a different notation for the specializations:
\beq
\tilde{Z}_{\rm pB}(n,p) = Z_{\rm pB}(n,p)|_{x_i=xq^{i-1}}, \qquad \tilde{E}_{(n;0,p)}={E}_{(n;0,p)} (x,qx,q^2x,\ldots,q^{n-1}x).
\eeq
Using~\eqref{SumSchurqx}, the expressions for $\tilde{E}_{(n;0,p)}$ again simplify a lot.
For our earlier example~\eqref{exE}, one finds
\beq
\tilde{E}_{(4;0,2)} = 1-q^3(1+q+q^2+q^3)x^3+q^6(1+q+q^2+q^3)x^5-q^{12} x^8 .
\label{exExq}
\eeq
and thus
\beq
\tilde{Z}_{\rm pB}(4,2)=\frac{1-q^3(1+q+q^2+q^3)x^3+q^6(1+q+q^2+q^3)x^5-q^{12} x^8}{\prod_{i=1}^4(1-xq^{i-1}) \prod_{1\leq j<k\leq 4}(1-x^2q^{j+k-2}) }.
\eeq

As before, let us also list some general examples:
\begin{itemize}
\item For $p\geq n$:
\begin{equation}
\tilde{Z}_{\rm pB}(n, p) = \frac{1}{\prod_{i=1}^n(1-xq^{i-1}) \prod_{j<k}(1-x^2q^{j+k-2}) }, 
\label{GPFxq}
\end{equation}
\begin{equation}
\bar{l}_i = \frac{xq^{i-1}}{1-xq^{i-1}}+ x^2\sum_{\myatop{\ss j=i-1}{\ss j\neq 2i-2}}^{n+i-2}\frac{q^j}{1-x^2q^j}.
\label{av_p>n-1xq}
\end{equation}
\item For $p=1$ (bosons):
\begin{equation}
\tilde{Z}_{\rm pB}(n, 1) = \frac{1}{\prod_{i=1}^n(1-xq^{i-1}) }, \qquad \bar{l}_i = \frac{xq^{i-1}}{1-xq^{i-1}}.
\label{av_p=1xq}
\end{equation}
\item For $p=n-1$:
\begin{equation}
\tilde{Z}_{\rm pB}(n, n-1) = \frac{1-x^nq^{n(n-1)/2}}{\prod_{i=1}^n(1-xq^{i-1})\prod_{j<k}(1-x^2q^{j+k-2}) }, \label{Z_p=n-1xq}
\end{equation}
\begin{equation}
\bar{l}_i = \frac{xq^{i-1}}{1-xq^{i-1}}+x^2\sum_{\myatop{\ss j=i-1}{\ss j\neq 2i-2}}^{n+i-2} \frac{q^j}{1-x^2q^j}
-\frac{x^nq^{n(n-1)/2}}{1-x^nq^{n(n-1)/2}}.
\label{av_p=n-1xq}
\end{equation}
\end{itemize}

Let us investigate the behavior of the average number of particles on orbital~$i$ as a function of the energy gap $\Delta$.
This is illustrated in Figure~2, where the ``population'' of the orbitals $\bar{l}_i$ ($p=5$, $n=5$ 
and $i=1, 2, 3, 4, 5$) is given as a function $q=\exp(-\Delta/\tau)$
(and for a chosen fixed value of $x$; here $x=\exp(-0.5)$).
Clearly, as $i$ increases, $\bar{l}_i$ decreases. 
So, the ``population'' of the orbitals depends on their level $i$. 
If $q=\exp(-\Delta/\tau)<<1$, i.e.\ for large gaps between the energy levels or for very low temperature, 
the particles are primarily found on the first orbital, i.e.\ the one with the lowest energy. 
For any value of $q$ ($0 < q < 1$) one has  $\bar{l}_1 >\bar{l}_2 >\bar{l}_3 >\bar{l}_4 >\bar{l}_5$.
For small values of $q$, $\bar{l}_1$ is large and the other average occupation numbers close to zero. 
The averages on the other orbitals become larger when $q$ increases.

\section{Thermodynamic properties of parafermion systems}
\setcounter{equation}{0} \label{sec:C}

The study of the GPF and of thermodynamic properties of parafermion statistics is similar, and we shall give less details than in the paraboson case.
Consider an $m$-parafermion system of order~$p$~\eqref{n-parafermion}-\eqref{pFockF}. 
For such systems, the GPF is of type~\eqref{GPF_general} with the sum restricted to partitions $\lambda$ 
for which $\lambda_1\leq p$~\cite{Chaturvedi96, Chaturvedi97a, Chaturvedi97b}, namely 
\begin{equation}
Z_{\rm pF}(m, p) =  \sum_{\lambda,\ \ell(\lambda')\leq p}  s_\lambda (x_1, x_2, \cdots, x_m) \label{pF_GPF}.
\end{equation}
Contrary to the paraboson case, this is a finite sum, so the GPF is already in proper form.
However, it will be useful to develop an alternative expression for $Z_{\rm pF}(m, p)$.
Since the GPF is again equal to the character of the corresponding Fock space $W(p)$, we can use the character formula 
given in~\cite{SV2015} or~\cite{King2013}. One finds for any positive integer~$p$,
\begin{equation}
Z_{\rm pF}(m, p) = \frac{E_{(m;p,0)}}{\prod_i(1-x_i) \prod_{j<k}(1-x_jx_k) },
\label{pF_GPFp}
\end{equation}
where 
\begin{equation}
E_{(m;p,0)} = \sum_{\mu} (-1)^{d_\mu} s_\mu (\boldsymbol{x}),
\label{F}
\end{equation}
with the sum  over all partitions $\mu$ of the form
\begin{equation}
\mu = \left(
\begin{array}{cccc}
b_1+p & b_2+p & \cdots & b_r+p \\
b_1 & b_2 & \cdots & b_r
\end{array} \right)
\label{mu}
\end{equation}
in Frobenius notation, and 
\begin{equation}
d_\mu= b_1+b_2+\cdots+b_r+r.
\label{d_mu}
\end{equation}

Without going into the details of the theory, let us mention that for $p=1$ the expression~\eqref{F} simplifies,
\begin{equation}
E_{(m;1,0)} =
\prod_{i=1}^m(1-x_i^2)\prod_{1\leq j<k\leq m}(1-x_j x_k) 
\end{equation}
and thus
\begin{equation}
Z_{\rm pF}(m, 1) =  \prod_{i=1}^m(1+x_i),
 \label{pF_GPFp=1}
\end{equation}
which is the GPF for an ordinary $m$-fermion system.

Let us next turn to the thermodynamic functions, the average number of particles $\bar{l}_i$ on the $i$th orbital 
and the average number of particles $\bar{N}$ in the system.
The general formulas follow from~\eqref{li}--\eqref{av}, applied to the current GPF~\eqref{pF_GPF} or~\eqref{pF_GPFp}
and will not be displayed in detail.

To describe some properties of these thermodynamic functions, let us as before consider two special situations.

First, consider the $m$-parafermion system of order~$p$ with identical energy levels.
As in subsection~\ref{3.2}, all orbitals have the same energy and the same chemical potential, thus $x_1=\cdots=x_m=x$.
Under this specialization, the GPF and ${E}_{(m;p,0)}$ will be denoted by 
$\breve{Z}$ and $\breve{E}_{(m;p,0)}$:
\beq
\breve{Z}_{\rm pF}(m,p) = Z_{\rm pF}(m,p)|_{x_i=x}, \qquad \breve{E}_{(m;p,0)}={E}_{(m;p,0)} (x,x,\ldots,x).
\eeq
When all $x_i$'s are equal, the distribution $\bar{l}_i$ is independent of $i$.
As a generic example, we plot the distribution functions $\bar{l}_i$ for the case $m=4$ and for $p=1, 2, 3, 4, 5$ in Figure~3 
(as before, the variable is $y$ as in~\eqref{y}).
Note that the average number of particles on the $i$th orbital, $\bar{l}_i$, does not exceed $p$ -- the order of the statistics,
which confirms the Pauli principle for parafermion statistics. 
For $p=1$, the distribution function coincides with the Fermi-Dirac distribution. 

Secondly, consider the $m$-parafermion system of order~$p$ with equidistant energy levels for the $m$ orbitals.
Following the notation of subsection~\ref{3.3}, the energy gap is $\Delta$, the orbital energies 
$\epsilon_i=\epsilon_1+(i-1)\Delta$, $\mu_1=\mu_2=\cdots=\mu_m=\mu$ and $x_i=x q^{i-1}$ with $x$ and $q$ given
in~\eqref{xq}.
We will use the following notation for the specializations:
\beq
\tilde{Z}_{\rm pF}(m,p) = Z_{\rm pF}(m,p)|_{x_i=xq^{i-1}}, \qquad \tilde{E}_{(m;p,0)}={E}_{(m;p,0)} (x,qx,q^2x,\ldots,q^{m-1}x).
\eeq
The behavior of the average number of particles on orbital~$i$ as a function of the energy gap $\Delta$ is illustrated in Figure~4.
We plot again the ``population'' of the orbitals $\bar{l}_i$ ($p=5$, $m=5$ 
and $i=1, 2, 3, 4, 5$) as a function $q=\exp(-\Delta/\tau)$
(and for a fixed value $x=\exp(-0.5)$).
The phenomena are comparable to the paraboson case with equidistant energy levels.
For any value of $q$ ($0 < q < 1$) one has  $\bar{l}_1 >\bar{l}_2 >\bar{l}_3 >\bar{l}_4 >\bar{l}_5$.
For small values of $q$, $\bar{l}_1$ is large and the other average occupation numbers small. 
The averages on the other orbitals become larger when $q$ increases.

\section{Remarkable properties of GPF's for paraboson and parafermion systems}
\setcounter{equation}{0} \label{sec:D}

In the case of identical energy levels there is a significant relation between the GPF of a paraboson system 
and the GPF of a parafermion system.
For this, notice that
\beq
\breve{E}_{(n;0,p)}= {E}_{(n;0,p)} (x,x,\ldots,x)= \sum_\eta (-1)^{c_\eta} s_\eta(x,x,\ldots,x),
\eeq
where the sum runs over partitions with Frobenius form~\eqref{eta}, whereas
\beq
\breve{E}_{(m;p,0)}= {E}_{(m;p,0)} (x,x,\ldots,x)= \sum_\mu (-1)^{d_\mu} s_\mu(x,x,\ldots,x),
\eeq
where the sum now runs over partitions with Frobenius form~\eqref{mu}.
The partitions appearing in these two sums are conjugate to each other, and from~\eqref{SumSchurx} it is
clear that
\beq
s_\lambda(x,\cdots,x) = s_{\lambda'} (x,x,\cdots,x),
\eeq
provided the number of $x$'s in the left hand side is $N+\ell(\lambda)$ and the number of $x$'s in the right hand side is $N+\ell(\lambda')$
(for any non-negative integer~$N$).
Using furthermore $s_\lambda(x_1,x_2,\ldots,x_k)=0$ for $\ell(\lambda)>k$, one deduces:
\beq
\breve{E}_{(n;0,p)}=\breve{E}_{(n+p;p,0)}.
\eeq
Since
\begin{align}
& \breve{Z}_{\rm pB}(n, p) = \frac{\breve{E}_{(n;0,p)}}{(1-x)^n (1-x^2)^{n(n-1)/2} }, \\
& \breve{Z}_{\rm pF}(m, p) = \frac{\breve{E}_{(m;p,0)}}{(1-x)^m (1-x^2)^{m(m-1)/2} } , 
\end{align}
one finds the following remarkable relation:
\beq
\breve{Z}_{\rm pB}(n, p) = \frac{\breve{Z}_{\rm pF}(n-p, p)}{(1-x)^p (1-x^2)^{np-p(p+1)/2} },
\qquad\hbox{for } p=1,2,\ldots,n
\eeq
($p=n$ can formally be included as ${Z}_{\rm pF}(0, p)=1$).
This is a surprising result: 
the GPF for a paraboson system can be computed from that of a parafermion system (in the case of identical energy levels).
The same then holds for the thermodynamic functions in this situation.

Let us next turn to some special values for $p$.
For $p=2$ one computes, e.g.\ from~\eqref{pF_GPF}, using~\eqref{SumSchurx}:
\begin{align*}
& \breve{Z}_{\rm pF}(1, 2) = 1+x+x^2, \\
& \breve{Z}_{\rm pF}(2, 2) = 1+2x+4x^2+2x^3+x^4, \\
& \breve{Z}_{\rm pF}(3, 2) = 1+3x+9x^2+9x^3+9x^4+3x^5+x^6, \\
& \breve{Z}_{\rm pF}(4, 2) = 1+4x+16x^2+24x^3+36x^4+24x^5+16x^6+4x^7+x^8, \\
& \breve{Z}_{\rm pF}(5, 2) = 1+5x+25x^2+50x^3+100x^4+100x^5+100x^6+50x^7+25x^8+5x^9+x^{10}.
\end{align*}
It is not difficult to see that there is a general formula for $p=2$, although it falls outside the scope of this paper to prove this:
\beq
\breve{Z}_{\rm pF}(m, 2) = \sum_{k=1}^m {m \choose k}^2 x^{2k} + \sum_{k=1}^{m} {m \choose k}{m \choose k-1} x^{2k-1}.
\eeq
We have obtained a similar general result for $p=3$, which we just mention here:
\beq
\breve{Z}_{\rm pF}(m, 3) = \frac{1}{m+2}(1+x)^m\left(\sum_{k=0}^m {m \choose k} {m+2 \choose k+1} x^{2k}\right).
\eeq
So far, we have not been able to find such alternative expressions for $p\geq 4$.
Note, by the way, that the above formulas are confirmed by substituting $x=1$ and comparing with the dimension formulas for the $m$-parafermion Fock spaces~\cite[eq.~(3.11)]{parafermion}:
\beq
\dim W(2) = \frac12 {2m+2\choose m+1}, \qquad \dim W(3) = \frac{2^m}{m+2} {2m+2 \choose m+1}.
\eeq

\section{Conclusions}
\setcounter{equation}{0} \label{sec:E}

Parastatistics was proposed many years ago as an extension of ordinary Bose-Einstein or Fermi-Dirac statistics, 
and depends on the introduction of an integer parameter~$p$ called the order of statistics.
Paraboson and parafermion systems of order $p=1$ just coincide with boson and fermion systems, 
but for $p>1$ the behaviour of the systems is different.

Parastatistics was studied by many people, but a complete account of statistical and thermodynamic properties 
of paraboson and parafermion systems of order~$p$ was never given (even though many special cases have been published).
The reason for this is that a general grand partition function was not known, 
at least not in a form that can be used to compute such properties.

Over the last couple of years, there have been many mathematical developments that shed a new light on this topic.
In particular, new forms of the character of paraboson and parafermion Fock spaces were obtained~\cite{paraboson,parafermion,SV2015}.
This allows the presentation of new and interesting expressions for the corresponding grand partition functions, given in this paper.
Using these grand partition functions for $n$-paraboson systems and $m$-parafermion systems of order~$p$, 
we have computed and discussed some thermodynamic properties of these systems, 
such as the average number of particles and orbital distributions.

Two specific cases have been considered in further detail. 
The case with identical energy levels per orbital leads to simplifications of the thermodynamic functions, 
and the corresponding distribution functions are reminiscent of the Bose-Einstein (for paraboson systems) 
and Fermi–Dirac (for parafermion systems) distributions.
But there are also some striking differences, as illustrated e.g.\ in Figures~1 and~3. 
The second specific case, with equidistant energy levels, also yields some interesting physical interpretations, 
e.g.\ concerning the orbital distribution (see Figures~2 and~4). 

One of our significant results is the relation between the GPF of $n$-paraboson systems of order $p$ 
and the GPF of $(n-p)$-parafermion systems of order $p$ (in the case of identical energy levels).
Such a relation could be established only due to the new forms of these GPF's.
Another outcome are the closed form expressions for these GPF's for $p=2$ and $p=3$.
We consider the last as an interesting result as paraparticles of order $p=2$ are recognized as candidates 
to be associated with dark matter and/or dark energy~\cite{Nelson}.

\section*{Acknowledgments}
N.I.\ Stoilova was supported by the Bulgarian National Science Fund, grant KP-06-N28/6, and J.\ Van der Jeugt was supported by the EOS Research Project 30889451.

\newpage
\begin{figure}
\caption{
Graphs of the average number of particles $\bar N(n,p)$ in terms of the variable $y=(\epsilon-\mu)/\tau$, 
for fixed $n=4$, and $p=1,2,3,4,5,\ldots$, for the $n$-paraboson system of order~$p$, in the case where the $n$ energy levels are equal.
The graph of $\bar N(4,1)$ is the closest to the $y$-axis, then $\bar N(4,2)$, etc.
The Bose-Einstein distribution coincides with $\bar N(4,1)$.
}
\[
\includegraphics{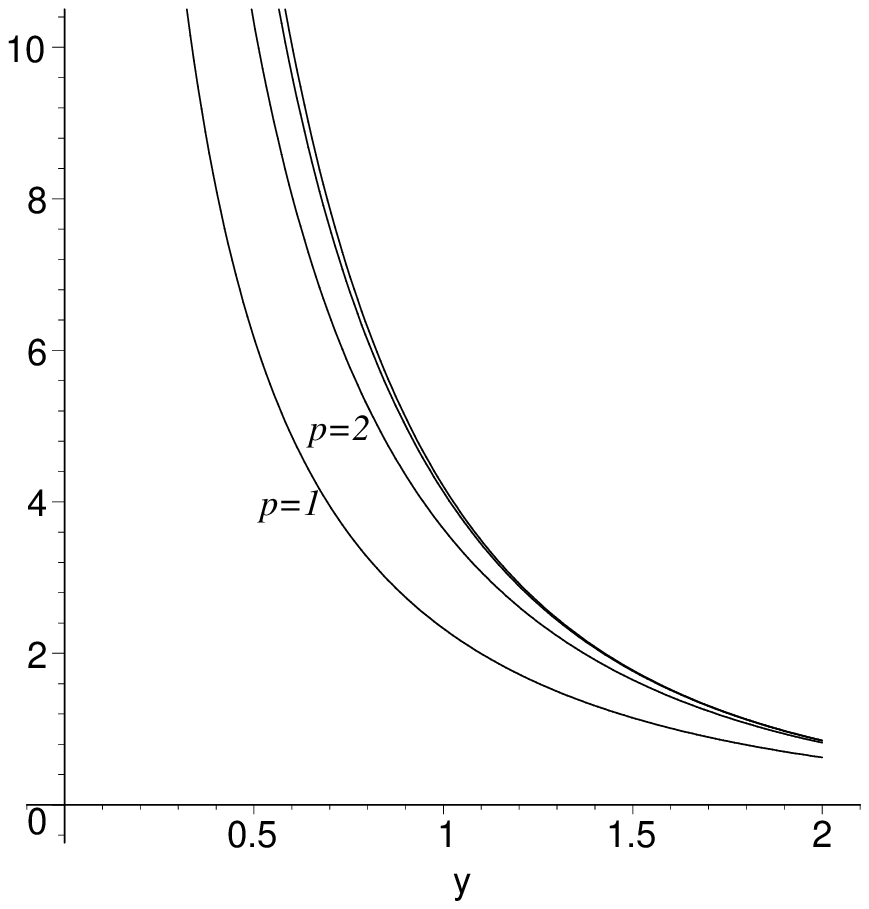} 
\]
\end{figure}

\begin{figure}
\caption{
Graphs of the average number of particles $\bar{l}_i$ on orbital~$i$, as a function of $q$ (i.e.\ depending on the energy gap).
The plot is for an $n$-paraboson system of order~$p$, with $n=5$ and $p=5$.
The graph of $\bar{l}_i$ is indicated by $i$ ($i=1,2,3,4,5$).
}
\[
\includegraphics{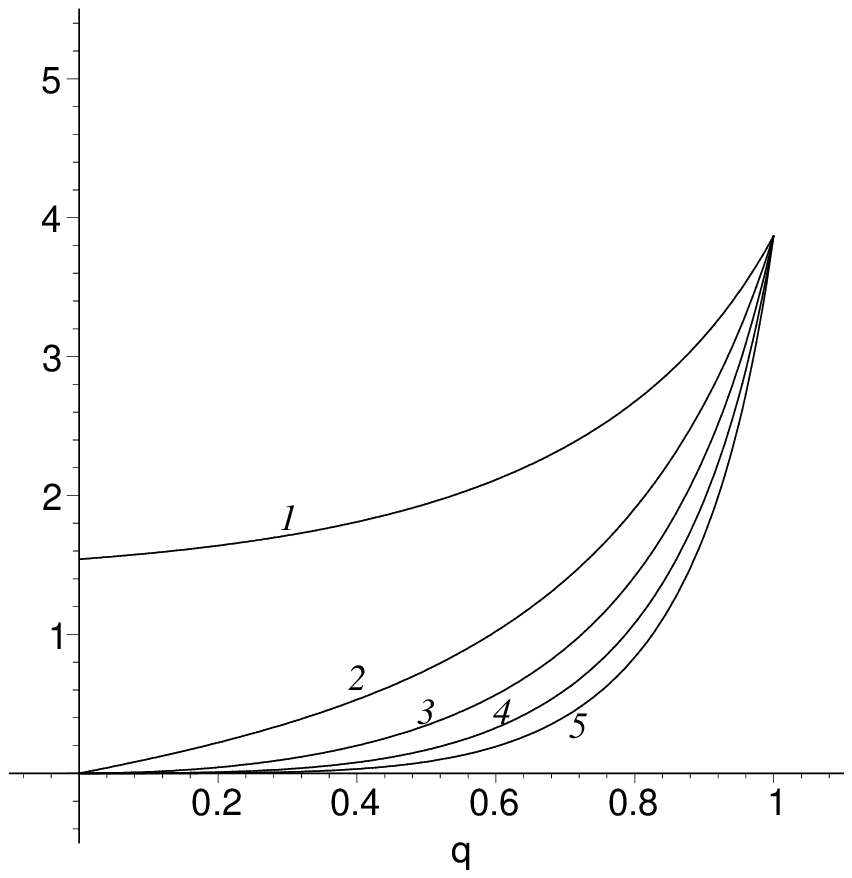} 
\]
\end{figure}

\begin{figure}
\caption{
Graphs of the average number of particles $\bar{l}_i$ on orbital~$i$, in terms of the variable $y=(\epsilon-\mu)/\tau$,
for fixed $m=4$, and $p=1,2,3,4,5$, for the $m$-parafermion system of order~$p$, in the case where the $m$ energy levels are equal
(and thus $\bar{l}_i$ does not depend on $i$).
The Fermi-Dirac distribution coincides with $p=1$.
}
\[
\includegraphics{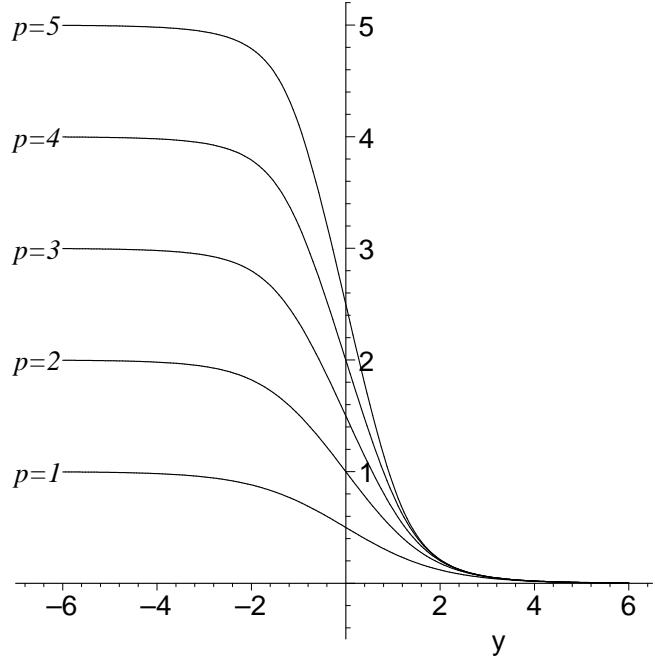} 
\]
\end{figure}

\begin{figure}
\caption{
Graphs of the average number of particles $\bar{l}_i$ on orbital~$i$, as a function of $q$ (i.e.\ depending on the energy gap).
The plot is for an $m$-parafermion system of order~$p$, with $m=5$ and $p=5$.
The graph of $\bar{l}_i$ is indicated by $i$ ($i=1,2,3,4,5$).
}
\[
\includegraphics{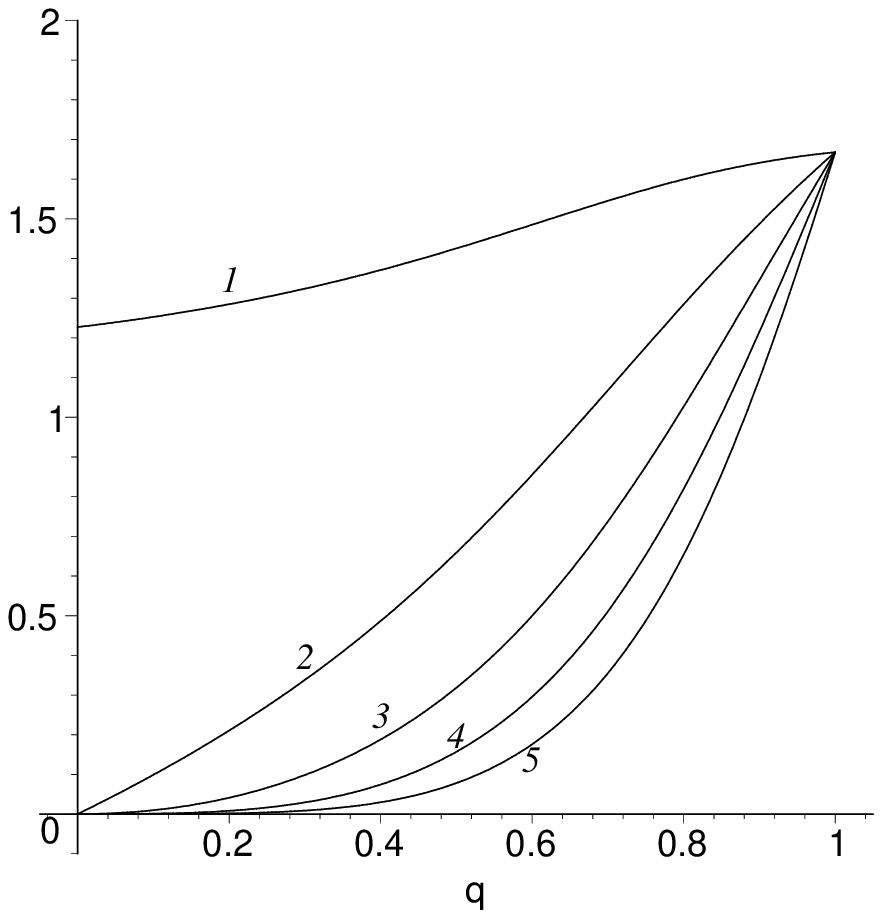} 
\]
\end{figure}

\end{document}